\documentclass[11pt, a4paper]{article}
\usepackage[utf8]{inputenc}
\usepackage{subcaption}
\usepackage{float}
\usepackage{lineno}
\usepackage[margin=3cm]{geometry}
\usepackage{graphicx}
\usepackage{amsmath}
\usepackage{amsfonts}
\usepackage{amssymb}
\usepackage[colorlinks=true
,urlcolor=blue
,anchorcolor=blue
,citecolor=blue
,filecolor=blue
,linkcolor=blue
,menucolor=blue
,pagecolor=blue
,linktocpage=true
,pdfa=true
]{hyperref}
\usepackage{wrapfig}
\usepackage[numbers,sort&compress]{natbib}
\usepackage{authblk}

\title{Instrumentation Challenges of the Strong-Field QED Experiment LUXE at the European XFEL}

\author{K. Fleck\thanks{kfleck02@qub.ac.uk}\\
On behalf of the LUXE collaboration.}
\affil{\textit{School of Mathematics and Physics,}\\\textit{The Queen's University of Belfast,}\\ \textit{Belfast, United Kingdom, BT7 1NN}}

\date{}

\begin{document}

\maketitle

\begin{abstract}
The LUXE experiment aims at studying strong-field QED in electron-laser and photon-laser interactions, with the $16.5$ GeV electron beam of the European XFEL and a laser beam with power of up to $350$ TW.  The strong-field QED processes are expected to have production rates ranging from $10^{-3}$ to $10^9$ per $10$ Hz bunch crossing. Additionally, these measurements must be performed in a low-energy, high-radiation background. The LUXE experiment will utilise various detector technologies to overcome these challenges.
\end{abstract}

\section{Introduction}
At the Laser Und XFEL Experiment (LUXE), a collaboration of multiple institutions including DESY Hamburg and the European XFEL (EuXFEL) \cite{Abramowicz2021}, it is envisioned to use the high quality $16.5$ GeV electron beam of the EuXFEL alongside a high power laser, up to $350$ TW, to investigate not only the transition from the perturbative to non-perturbative regime of QED, but also farther into the less studied strong field regime \cite{Bula:1996,Burke:1997}.
Three signature processes of QED are to be measured at LUXE: non-linear Compton scattering ($e^{\pm} + n\gamma_L \rightarrow e^{\pm} + \gamma$); non-linear Breit-Wheeler pair production \cite{Breit:1934} ($\gamma + n\gamma_L\rightarrow e^+e^-$) and the non-linear trident process \cite{ritus72} ($e^{\pm} \rightarrow e^{\pm} + e^+e^-$). $n\gamma_L$ denotes the interaction with multiple laser photons in an instantaneous process. To this end, two operating modes are proposed; the electron-laser mode (direct interaction of the electron beam and the laser); and gamma-laser mode (electron beam converted to bremsstrahlung before interaction). \par
\begin{figure}[H]
    \centering
    \begin{subfigure}[b]{0.3\textwidth}
        \centering
        \includegraphics[width=\textwidth]{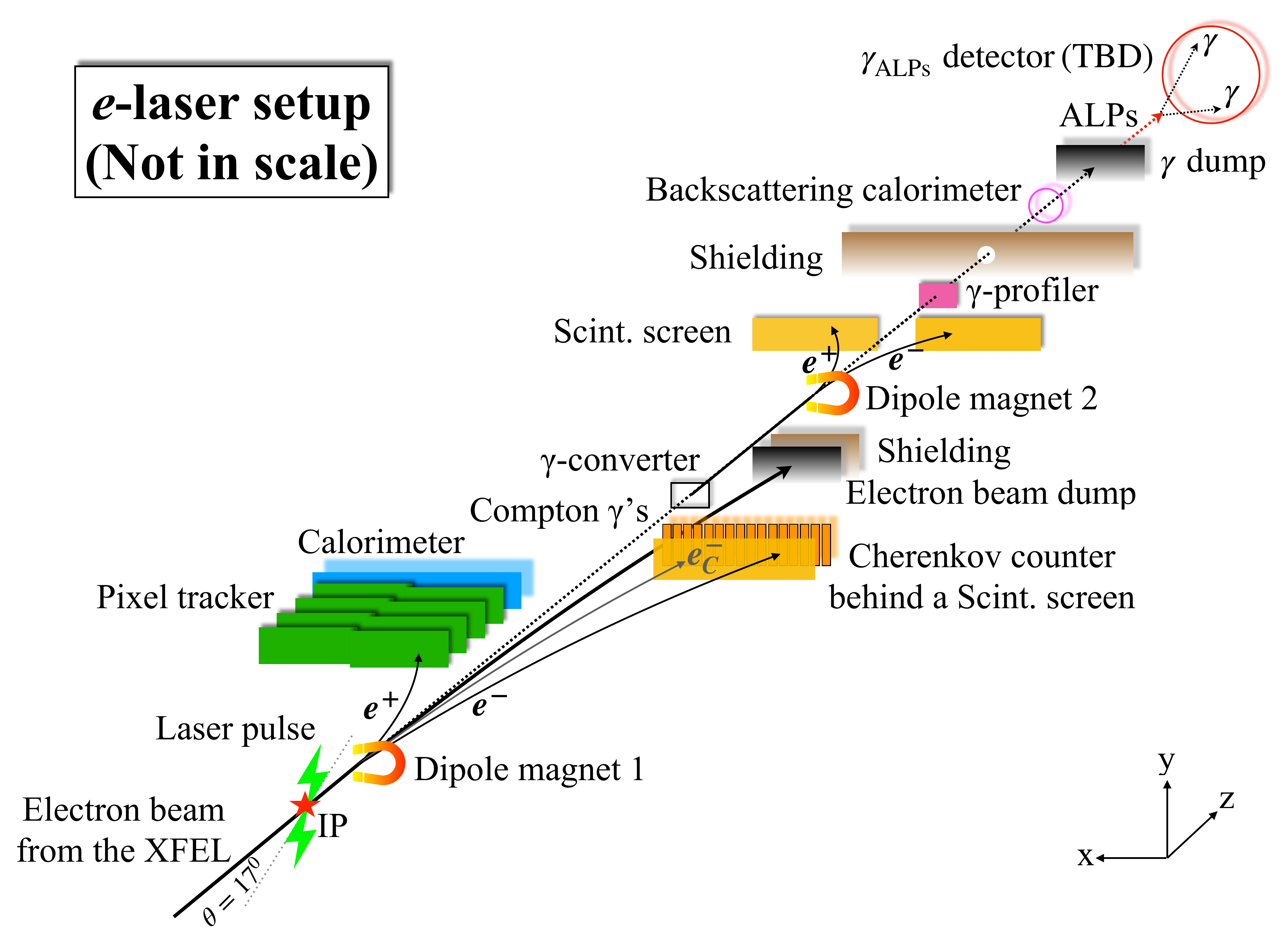}
        \label{fig1.a}
        \caption{}
    \end{subfigure}%
    \hspace{10mm}
    \begin{subfigure}[b]{0.3\textwidth}
        \centering
        \includegraphics[width=\textwidth]{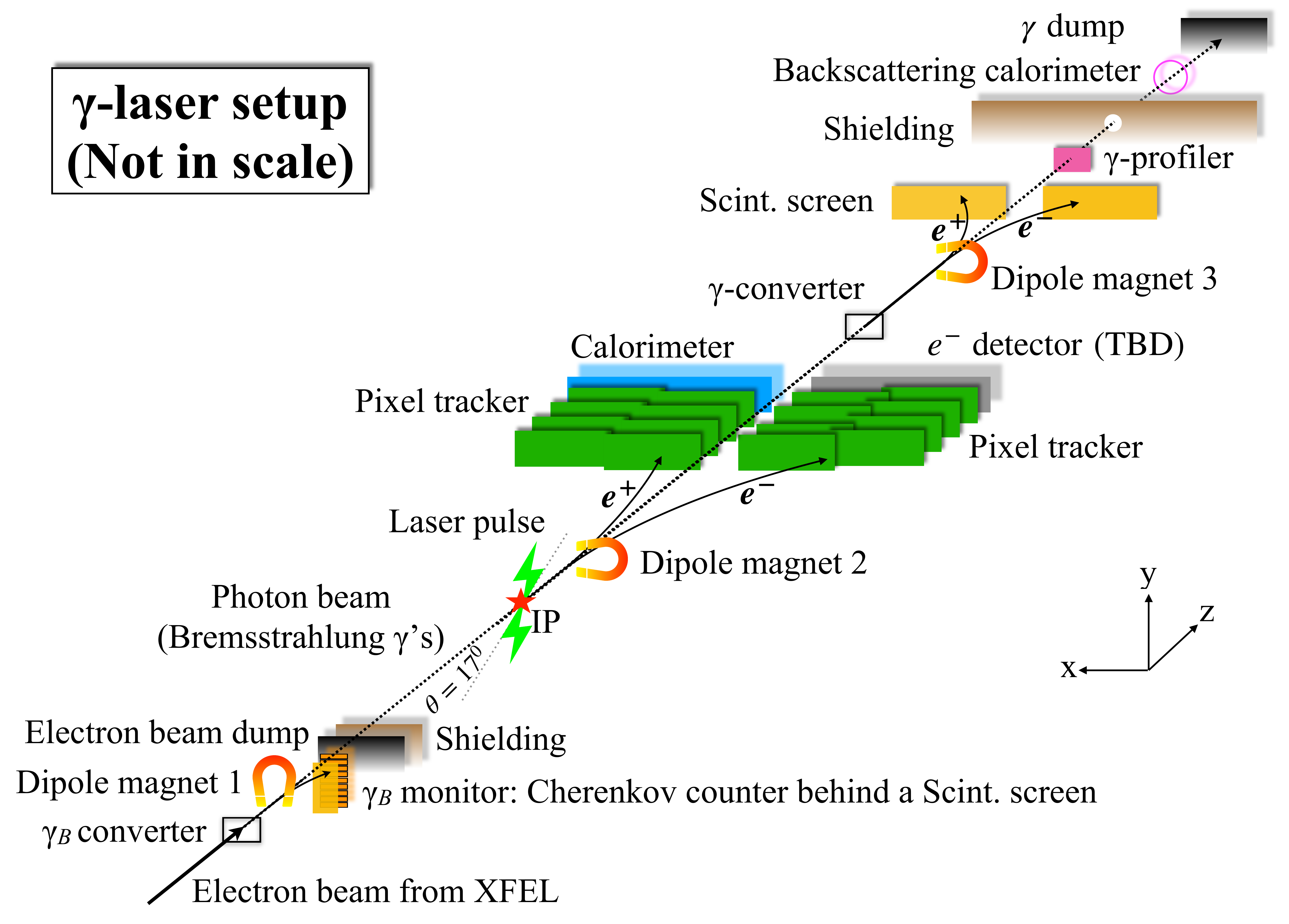}
        \label{fig1.b}
        \caption{}
    \end{subfigure}
    \caption{Schematic of the setup for each operating mode at LUXE, (a) the electron-laser mode, and (b) the gamma-laser mode. Relative locations of each detector subsystem are shown, along with additional magnets, beam dumps and shielding.}
    \label{fig:1}
\end{figure}

Measurement of these processes requires essentially three detector systems for the anticipated electrons, positrons and photons. Due to the nature of the production processes, the signal expected across each system varies over many orders of magnitude, from $10^{-4}-10^9$ per 1 Hz laser bunch crossing (BX), depending on operating mode and particle type. Therefore, it is paramount that the detectors involved are able to resolve signal from similar level but lower energy background across this range of magnitudes.

\section{Electron detection}
Detection of electrons at LUXE presents a challenge due to their high flux (up to $10^9$ per BX) and high combined energy, so two complementary technologies have been proposed to ensure both efficient collection of signal and its resolution from background. The electron detection system is situated downstream of the interaction point (IP), after a dipole magnet, allowing magnetic spectroscopy methods to be used. \par 

\begin{figure}
    \centering
    \begin{subfigure}[b]{0.3\textwidth}
        \centering
        \includegraphics[width=\textwidth]{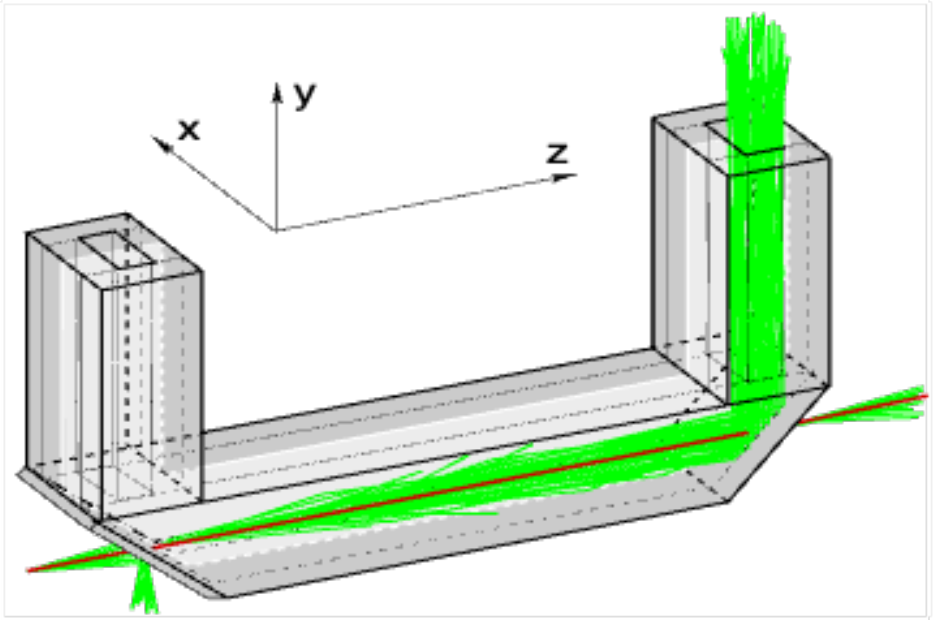}
        \caption{}
        \label{fig:2a}
    \end{subfigure}%
    \begin{subfigure}[b]{0.3\textwidth}
        \centering
        \includegraphics[width=\textwidth]{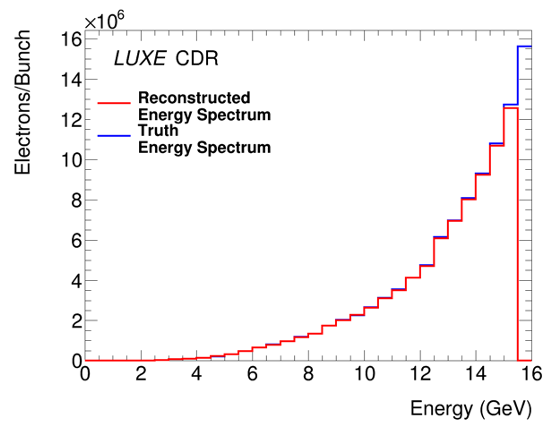}
        \caption{}
        \label{fig:2b}
    \end{subfigure}
    \caption{(a) Schematic of a single U-shaped \u{C}erenkov channel with an incident electron marked in red and the generated optical light in green. The reflection of the light by a low transmission mirror is also depicted. (b) Simulated reconstruction of the electron energy spectrum from the scintillator screen, with the blue showing the true spectrum and red, the reconstruction. The final bin is zero in the reconstruction by design.}
    \label{fig:2}
\end{figure}

Following designs delevoped for polarimetry experiments at future lepton colliders \cite{Bartels:2010}, the \u{C}erenkov detector is composed of U-shaped aluminium channels filled with an active gas medium with low refractive index as in Figure \ref{fig:2a}. As the energy separated electrons pass through a channel, \u{C}erenkov radiation is emitted and reflected using mirrors into an imaging camera fitted with optical filters for protection. Robust against high particle fluxes, the \u{C}erenkov detector is ideal for measurement of the high flux (and higher energy) portion of the electron signal while having lower sensitivity to low energy noise ($E < 20$ MeV) and no response to any photon background. From simulations in GEANT4 \cite{Agostinelli:2002,Allison:2006,Allison:2016}, it has been possible to reconstruct the energy spectrum using the spatially resolved information to within $0.3 - 0.5$ \%.\par

The scintillator screen, chosen to be terbium-doped gadolinium oxysulfide (Gd$_2$O$_2$S:Te), also provides high resolution in the position detection of the electron signal. Additionally, the disturbance of the electron signal will be minimal and so can be combined with the \u{C}erenkov technology effectively. Figure \ref{fig:2b} shows the reconstruction of the electron energy spectrum from simulated position measurements of the scintillator.

\section{Positron detection}
In the photon-laser mode, the positron detection system expects the same level of signal as the electron side, however in the electron-laser mode, the positron rate varies greatly from $10^{-4} - 10^7$ per BX. This challenge of detecting a signal with such a large range of flux is met by employing a tracking detector and a calorimeter. Together, they allow for high-granularity measurement of the signal at low flux and accurate measurement of particle numbers at high fluxes.\par
The tracking subsystem is comprised of 8 tracker staves - each stave is an approximately $27$ cm long frame containing appropriate cooling and a hybrid integrated circuit (HIC), based on the ALPIDE silicon pixel detector \cite{Mager:2016} to be used at ALICE, LHC. The staves are arranged in four layers to form the tracking system. A Kalman filter fitting algorithm \cite{Billoir:1990} is used to reconstruct the signal passing through all four layers, while rejecting background based on energy, geometrical and statistical constraints; this has been shown to give a $>95\%$ reconstruction efficiency for $E\geq 2.5$ GeV. This analysis was performed using simulation data from GEANT4 of the expected signal and background at the tracking subsystem. Further reconstruction of the positron energy spectrum is predicted to be possible to within $1 \%$.\par

In conjunction with the tracking detector, an electromagnetic calorimeter, following the design of the FCAL collaboration \cite{Abramowicz:2018}, is to be used to give an independent measurement of the positron flux and their energy spectrum. The active part of the calorimeter is composed  is composed of 20 layers of tungsten, interspersed with $1$ mm gaps containing silicon sensor plates. Due to the low Moli\`ere radius of the detector, this allows for good resolution of local energy deposits. From simulations, the position resolution of positrons is $780\, \mu$m, leading to an achievable resolution in the positron energy spectrum of $\sim 20\%$ at $1$ GeV, improving with increasing positron energy.

\section{Photon detection and monitoring}
The gamma ray spectrometer, as discussed in Ref. \cite{Fleck:2020}, is one of the three technologies to be used in the photon detection system. A pair of scintillator screens measure the spatially resolved $e^+e^-$ pairs produced by passing the photon beam through a thin converter target. Again, due to the high photon flux, PMTs are not required for collection of the signal by a camera. Simulations have been performed to optimise the signal-to-background ratio ($\sim10-100$) and the energy resolution available at the detector ($< 1\%$ at $16.5$ GeV, and improves with decreasing energy). The deconvolution algorithm which reconstructs the original photon spectrum is currently being tested on data from another QED experiment. \par

\begin{figure}
    \centering
    \begin{subfigure}[b]{0.3\textwidth}
        \centering
        \includegraphics[width=\textwidth]{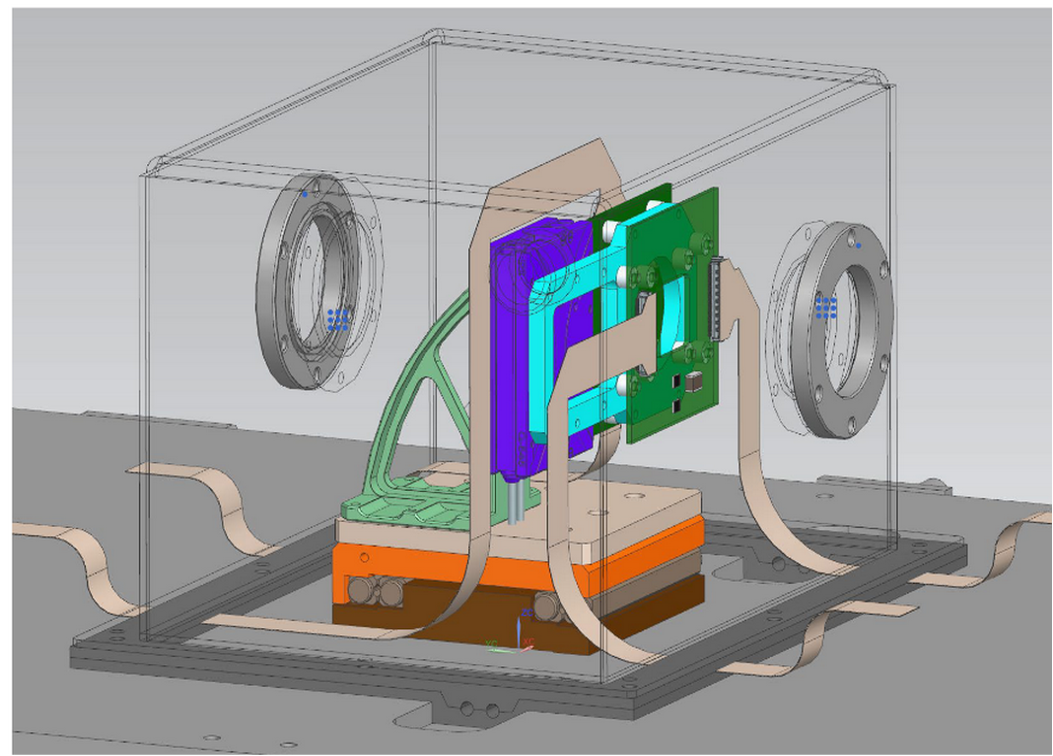}
        \caption{}
        \label{fig:3a}
    \end{subfigure}%
    \hspace{10mm}
    \begin{subfigure}[b]{0.3\textwidth}
        \centering
        \includegraphics[width=\textwidth]{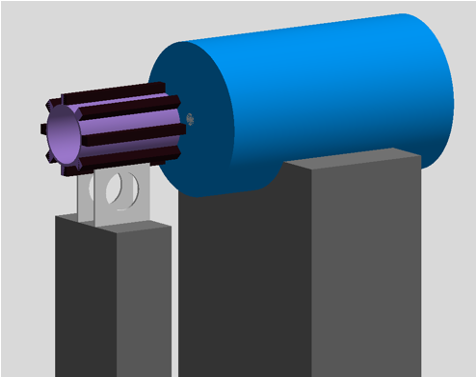}
        \caption{}
        \label{fig:3b}
    \end{subfigure}
    \caption{(a) CAD model of the current design of the GBP. (b) GEANT4 rendering of the GFM (left) and the photon beam dump (right, blue). }
    \label{fig:3}
\end{figure}

 The gamma ray profiler (GBP), as shown in Figure \ref{fig:3a}, aims to provide an in-situ diagnostic of the parameter $\xi$ using a technology similar to standard silicon strip detectors to measure the spatial profile of the photon beam. Instead of silicon, sapphire is to be used based on its preferred properties, such as its higher bandgap and good radiation hardness. Two sapphire planes, with mutually orthogonally oriented strips of pitch $\sim 100\,\mu$m, will be placed directly in the photon beam line in order to measure the photon number distribution in two orthogonal directions in a plane perpendicular to the photon propagation. After approximately 4 BXs, the spatial resolution will be at the $5 \, \mu$m level, using an analogue readout with a CoG algorithm, which is required to achieve a precision in $\xi$ of $2.5\%$.\par

The third component of the photon detection system is the gamma flux monitor (GFM). As the photon flux expected is high, particularly in the electron-laser mode, the GFM measures the flux indirectly using backscattering from the downstream beam dump. The GFM functions as a calorimeter, composed of eight lead-glass blocks, oriented parallel to the beam propagation axis, spaced on an open, cylindrical structure as shown in Figure \ref{fig:3b}. Upstream-pointing PMTs are used to collect the energy deposition information in the GFM. GEANT4 simulations of the GFM have shown that the energy deposited is directly proportional to the mean number of photons impinging the beam dump. As such, this can be calibrated using standard candles to give the photon number per BX with an uncertainty of $5-10\%$. These candles can also be used to monitor the stability of the lead-glass blocks in regards to radiation damage and deterioration.

\section{Conclusions}
LUXE aims to investigate new regimes of SFQED by means of carefully developed systems of detectors. These detectors are capable of distinguishing signal over a wide range of magnitudes ($10^{-4}-10^9$ per BX) from a high-flux, low-energy background.  Across all subsystems, a resolution of $\lesssim15\%$ in the quantity of interest has been calculated over the energy range of interest for the experiment. Furthermore, many of the technologies developed for use at LUXE can be adapted for use in other SFQED experiments and ought to provide a benchmark for the detector requirements of future investigations into non-perturbative QED.

\section{Acknowledgements}
The work presented in this document has been partially supported by the Engineering and Physical Sciences Research Council (grant No.: EP/T021659/1 and EP/V049186/1).

\bibliographystyle{unsrt}
\bibliography{References}

\end{document}